\documentstyle[12pt]{article}
\def\D{\hbox{D\kern-.73em\raise.25ex\hbox{-}\raise-.25ex\hbox{ }}}
 \def\d{\hbox{d\kern-.33em\raise.75ex\hbox{-}\raise-.75ex\hbox{}}}

\def\R{{\bf R}}

\def\GGG{\frak G }
\def\gr3{\GGG\,(\SSS_3)}

\def\gr2{\GGG\,(\SSS_2)}

\def\SSS{\frak S}

\def\al{{\alpha}}
\def\bet{{\beta}}

\def\gam{{\gamma}}

\def\l{\lambda }
\def\ve{\varepsilon }

\def\vp{\vspace}
\def\hp{\hspace}
\def\ed{\end{document}}
\def\beq{\begin{equation}}
\def\eeq{\end{equation}}
\def\bea{\begin{eqnarray}}
\def\eea{\end{eqnarray}}
\def\ba{\begin{array}}
\def\ea{\end{array}}
\def\bi{\begin{itemize}}
\def\ei{\end{itemize}}

\def\lb{\label}

\def\noi{\noindent}
\def\nn{\nonumber}

\begin{document}
\title{\bf Lagrangian Aspects of Quantum Dynamics on a Noncommutative Space
\thanks{{\footnotesize{Talk at the  Workshop {\it Contemporary Geometry
and Related Topics}, Belgrade, Yugoslavia, May 15-21, 2002. }}}
 }
\author{Branko Dragovich\thanks{e-mail address: dragovich@phy.bg.ac.yu}
\\
Institute of Physics, P.O. Box 57, 11001 Belgrade, Yugoslavia
\\
 Zoran Raki\'c\thanks{e-mail address: zrakic@matf.bg.ac.yu}
\\
Faculty of Mathematics, P.O. Box 550, 11000 Belgrade, Yugoslavia}
\date{}

\maketitle
\begin{abstract}
\noi In order to evaluate the Feynman path integral in
noncommutative quantum mechanics, we consider properties of a
Lagrangian  related to a quadratic Hamiltonian with noncommutative
spatial coordinates. A quantum-mechanical system with
noncommutative coordinates is equivalent to another one with
commutative coordinates. We found connection between quadratic
classical Lagrangians of these two systems. We also shown that
there is a subclass  of quadratic Lagrangians, which includes
harmonic oscillator  and particle in a constant field, whose
connection between ordinary and noncommutative regimes can be
expressed as a linear change of position in terms of a new
position and velocity.
\end{abstract}

\vspace{2mm}

 \noi MSC2000: 81T75, 81S40

 \noi PACS: 03.65.Bz

\vspace{1cm}


\section{Introduction}

Quantum theories with noncommuting spatial coordinates have been
investigated intensively during the recent years. M(atrix) theory
compactification on noncommutative tori, strings in some constant
 backgrounds, quantum Hall effect and IR/UV mixing are some of
the most popular themes (for a review of noncommutative quantum
field theory and some related topics, see e.g. \cite{1}). Most of
the research has been done in noncommutative field theory,
including noncommutative extension  of the Standard Model
\cite{2}. Since quantum mechanics can be regarded as the
one-particle nonrelativistic sector of quantum field theory, it is
also important to study its noncommutative aspects including
connection between ordinary and noncommutative regimes. Because of
possible phenomenological realization, noncommutative quantum
mechanics (NCQM) of a charged particle in the presence of a
constant magnetic field has been mainly considered on two- and
three-dimensional spaces (see, e.g. \cite{3} and references
therein). \vspace{1mm}

\noi Recall that to describe quantum-mechanical system
theoretically one uses a Hilbert space $L_2 (\R^n)$ in which
observables are linear self-adjoint operators. In ordinary quantum
mechanics (OQM), by quantization, classical canonical variables
$x_k,p_j$ become Hermitean operators $\hat{x_k},\hat{p_j}$
satisfying the Heinsenberg commutation relations \beq [
\hat{x_k},\hat{p_j}] = i\,\hbar\, \delta_{kj}, \quad
[\hat{x_k},\hat{x_j}] = 0,\quad  [\hat{p_k},\hat{p_j}] =0,\ \ \
k,j=1,2\dots n\, . \label{1} \eeq

\noi In a very general NCQM  one has that  $ [
\hat{x_k},\hat{p_j}] = i\,\hbar\, \delta_{kj}$, but also $
[\hat{x_k},\hat{x_j}] \neq 0$ and $[\hat{p_k},\hat{p_j}] \neq 0 .
$ However, we consider here the most simple and usual NCQM which
is based on the following algebra: \beq [ \hat{x_k},\hat{p_j}] =
i\,\hbar\, \delta_{kj}, \quad [\hat{x_k},\hat{x_j}] = i\,\hbar\,
\theta_{kj},\quad [\hat{p_k},\hat{p_j}] =0\, , \label{2} \eeq
where ${\bf\Theta}=(\theta_{kj})$ is the antisymmetric matrix with
constant elements. \vspace{1mm}

\noi To find $\Psi (x,t)$ as elements  of the Hilbert space in
OQM, and their time evolution, it is usually used the
Schr\"odinger equation
$$i\,\hbar\, \frac{\partial}{\partial t}\, \Psi (x,t) = \hat{H}\,\Psi
(x,t),$$ which realizes the eigenvalue problem for the
corresponding Hamiltonian operator $\hat{ H} = H (\hat{p}, x, t)$,
where $\hat{p_k} =-\,i\,\hbar\,\frac{\partial}{\partial x_k}$ and
$\hbar =\frac{h}{2\pi}$. However, there is another approach based
on the Feynman path integral method \cite{4} \beq {\cal
K}(x'',t'';x',t') =\int_{(x',t')}^{(x'',t'')} \exp \left (
\frac{i}{\hbar}\, S[q]\right )\, {\cal D}q , \label{3} \eeq where
${\cal K}(x'',t'';x',t')$ is the kernel of the unitary evolution
operator $U (t)$, functional $ S[q] = \int_{t'}^{t''} L(\dot{q},q,
t)\, dt$ is the action for a path $q(t)$ in the classical
Lagrangian $L(\dot{q},q,t)$, and $x''=q(t''), \ x'=q(t')$ with the
following notation $x =(x_1,x_2,\dots,x_n)$ and
$q=(q_1,q_2,\dots,q_n)$. The kernel ${\cal K}(x'',t'';x',t')$ is
also known as quantum-mechanical propagator,  Green's function,
and the probability amplitude for a quantum particle to come from
position $x'$ at time  $t'$ to another point $x''$ at $t''$. The
integral in (\ref{3}) has a symbolic meaning of an intuitive idea
that a quantum-mechanical particle may propagate from $x'$ to
$x''$ using infinitely many paths which connect these two points
and that one has to take all of them into account. Thus the
Feynman path integral means a continual (functional) summation of
single transition amplitudes $\exp \left ( \frac{i}{\hbar}\,
S[q]\right ) $ over all possible continual paths $q(t)$ connecting
$x'=q(t')$ and $x''=q(t'')$. In direct calculations, it is the
limit of an ordinary multiple integral over $N-1$ variables $q_i =
q(t_i)$ when $N\to \infty$. Namely, the time interval $t''-t'$ is
divided into $N$ equal subintervals and integration is performed
for every $q_i \in (-\infty,+\infty)$ and fixed time $t_i$. In
fact, ${\cal K}(x'',t'';x',t')$, as the kernel of the unitary
evolution operator, can be defined by equation \beq  \Psi
(x'',t'') = \int {\cal K}(x'',t'';x',t')\, \Psi (x't')\, dx'
\label{4} \eeq and then Feynman's path integral is a method to
calculate this propagator. Eigenfunctions of the integral equation
(\ref{4}) and of the above Schr\"odinger equation are the same.
Note that the Feynman path integral approach, not only to quantum
mechanics but also to whole quantum theory, is intuitively more
attractive and more transparent in its connection with classical
theory  than  the usual canonical operator formalism. In gauge
theories and some other cases, it is  also the most suitable
method of quantization. \vspace{1mm}

\noi The Feynman path integral for quadratic Lagrangians  can be
evaluated analytically and the exact result for the propagator is
\beq  {\cal K}(x'',t'';x',t') =\frac{1}{(i h)^{\frac{n}{2}}}
\sqrt{\det{\left(-\frac{\partial^2 {\bar S}}{\partial x''_k
\partial x'_j} \right)}} \exp \left(\frac{2\pi i}{h}\,{\bar
S}(x'',t'';x',t')\right), \label{5} \eeq where $ {\bar
S}(x'',t'';x',t')$ is the action for the classical trajectory
which is solution of the Euler-Lagrange equation of
motion.\vspace{1mm}

\noi In this article we search the form of a  Lagrangian which
corresponds to a system with noncommutative spatial coordinates.
This is necessary to know before to employ Feynman's path integral
method in NCQM. To this end, let us note that algebra (\ref{2}) of
operators $\hat{x_k}, \hat{p_j}$ can be replaced by the equivalent
one \beq [ \hat{q_k},\hat{p_j}] = i\,\hbar\, \delta_{kj}, \quad
[\hat{q_k},\hat{q_j}] = 0,\quad [\hat{p_k},\hat{p_j}] =0\, ,
\label{6} \eeq where linear transformation \beq \hat{ x_k} =
\hat{q_k} - \frac{\theta_{kj}\hat{p_j}}{2} \label{7} \eeq is used,
while $\hat p_k$ are remained unchanged, and summation over
repeated indices is assumed. According to (\ref{2}), (\ref{6}) and
(\ref{7}), NCQM related to  the classical phase space $(x,p)$ can
be regarded as an OQM on the other phase space $(q,p)$. Thus, in
$q$-representation, $\hat{p_k} = -\,i\,\hbar\,
\frac{\partial}{\partial q_k}$ in  the equations (\ref{6}) and
(\ref{7}). It is worth noting that Hamiltonians
$H(\hat{p},\hat{x}, t) =H(-i\hbar\frac{\partial}{\partial q_k},\,
q_k + \frac{i\hbar\theta_{kj}}{2}\frac{\partial}{\partial x_j},\,
t )$, which are more than quadratic in $\hat{x}$, will induce
Schr\"odinger equations which contain derivatives higher than
second order and even of the infinite order. This leads to a new
part of modern mathematical physics of partial differential
equations with arbitrary higher-order derivatives (for the case of
an infinite order, see, e.g. \cite{5} and \cite{6}). In this paper
we restrict our consideration to the case of quadratic
Langrangians in $x$ and $\dot{x}$ (for an example with 
$\ddot{x}$, see \cite{lukierski}).

\section{Quadratic Lagrangians}
\subsection{Classical case}
Let us start with a classical system described by a quadratic
Lagrangian which the most general form in three dimensions is:
\bea\nn \hspace{-5mm} L(\dot{x}, x,t) &=& \alpha
_{11}\,\dot{x}_1^2 +\alpha _{12}\,\dot x_1 \, \dot x_2  +\alpha
_{13}\,\dot x_1 \, \dot x_3 \, +\alpha _{22}\,\dot{x}_2^2 +\alpha
_{23}\,\dot x_2 \, \dot x_3 \,
  \vp{2mm} \\ & + & \alpha _{33}\,\dot{ x}_3^2 + \beta
_{11}\,{\dot x}_1 \, x_1  +\beta _{12}\, \dot x_1 \, x_2 +\beta
_{13}\,\dot x_1 \, x_3 + \beta _{21}\,\dot x_2 \, x_1\label{8}  \vp{2mm}\\
\nn & + &  \beta _{22}\,\dot x_2 \, x_2 +\beta _{23}\,\dot x_2 \,
x_3 + \beta _{31}\,\dot x_3\, x_1  +\beta _{32}\, \dot x_3
\, x_2 + \beta _{33}\, \dot x_3 \, x_3\vp{2mm}\\
 & + & \gamma _{11}\, x_1^{2}
+\gamma _{12}\, x_1 \, x_2 +\gamma _{13}\, x_1 \, x_3 +\gamma
_{22} \, x_2 ^{2} + \gamma _{23}\, x_2 \, x_3  + \gamma _{33}\,
x_3^{2} \vp{2mm} \nn \\
& + & \delta _1 \, \dot x_1  +\delta _2 \, \dot x_2 + \delta_3 \,
\dot x_3 + \xi_1 \, x_1 +\xi _2\, x_2 +\xi_3\, x_3 + \phi ,
\nn\eea where the coefficients $\alpha_{ij} =\alpha_{ij}(t),\
\beta_{ij} =\beta_{ij}(t),\ \gamma_{ij} =\gamma_{ij}(t),\
\delta_{i} =\delta_{i}(t),$ \ $\xi_{i} =\xi_{i}(t)$ and $\phi
=\phi(t)$ are some analytic functions of  the time $t$ .
\vspace{2mm}

\noi If we introduce the following matrices, \vp{5mm}

$
 \hspace{-8mm}\begin{array}{ccc}  {\alpha } = \left(
 \begin{array}{ccc}
 \alpha _{11} & {\alpha _{12} \over 2} &
{\alpha _{13} \over 2}\vspace{2mm} \\
{\alpha_{12} \over 2} & \alpha_{22} & {\alpha_{23} \over
2} \vspace{2mm}\\
 {\alpha_{13} \over 2 } & {\alpha_{23} \over 2} &
 \alpha_{33}
\end{array}
\right) ,   &   {\beta } = \left(
\begin{array}{ccc}
 \beta_{11} & \beta_{12} & \beta_{13} \\
 \beta_{21} & \beta_{22} & \beta_{23} \\
 \beta_{31} & \beta_{32} & \beta_{33} \end{array}
\right) ,  &  {\gamma }= \left(
\begin{array}{ccc} \gamma_{11} & {\gamma_{12} \over 2}
& {\gamma_{13} \over 2}\vspace{2mm} \\
{\gamma_{12} \over 2}  & \gamma_{22} & {\gamma_{23}\over 2}\vspace{2mm} \\
{\gamma_{13} \over 2 } & {\gamma_{23} \over 2} & \gamma_{33}
\end{array} \right) , \end{array}$ \vspace{4mm}

\noi assuming that the matrix $\al$ is nonsingular (regular)  and
if we introduce vectors \bea \nn {\delta } =(\delta _1 ,\delta_2
,\delta_3 ), \quad {\xi} = (\xi_1 , \xi_2 , \xi_3 ), \quad {\dot
x}= (\dot x_1, \dot x_2, \dot x_3), \quad {x}= (x_1 ,x
_2,x_3),\eea one can express the Lagrangian (\ref{8}) in the
following, more compact, form: \beq \label{lg} L(\dot{x},x,t)
=\langle\, {\alpha }\, {\dot x }, {\dot x } \,\rangle + \langle\,
{\beta }\, {x }, {\dot x }\, \textbf{}\rangle +\langle\, {\bf
\gamma }\, {x}, {x }\, \rangle + \langle\, {\delta } , {\dot x }\,
\rangle + \langle\, {\xi } , {x }\, \rangle + \bf \phi  ,
\end{equation}
where $\langle\, \cdot\, ,\, \cdot\, \rangle$ denotes standard
scalar product. Using the equations
\begin{equation}
p_j = {\partial L \over \partial \dot x_j}, \ \ \ j=1,2,3,
\end{equation}
one can express ${\dot x }$ as
\begin{equation}
{\dot x } = {1\over 2}\,\, {\alpha^{-1}}\, ({p}  - {\beta}\, {x }
- \delta ).
\end{equation}
Then, the corresponding classical  Hamiltonian \beq \label{clh}
H(p,x,t)=\langle\, {p}, \dot {x }\, \rangle - L(\dot{x},x,t)\eeq
becomes also quadratic, i.e., \beq \label{ch} H(p,x,t) =\langle\,
{A}\, {p}, {p}\, \rangle + \langle\, {B}\, {x}, {p }\, \rangle
+\langle\, {C}\, {x }, {x}\, \rangle + \langle\, {D } , {p}
\,\rangle + \langle\, { E } , {x} \rangle + F ,
\end{equation}
where: \bea
  & & {A } ={\frac 14}\,\,  {\alpha}^{-1}, \hspace{3cm}   {B }
=-\,{1\over 2} \,\, {\alpha}^{-1}\, {\beta}, \vp{2mm}\nn \\
& & {C} ={1\over 4}\,\, {\beta}^{\tau}\, {\alpha}^{-1}\, {\beta} -
{\gamma } , \hspace{1.37cm}
  {D} =- {1\over 2}\,\, {\alpha }^{-1}\,  \delta,\vp{2mm} \label{lcl}\\
& & {E } = {1\over 2}\,\, {\bf \beta}^{\tau} {\bf \alpha}^{-1}\,
{\bf \delta}  - {\xi} , \hspace{1.56cm} {F} = {1\over 4}\,\,
\langle\, {\delta} , {\alpha}^{-1} {\delta}\, \rangle - {\phi }
\nn \,. \eea  \noi Here,  $\beta^{\tau}$ denotes transpose map of
$\beta$.


\noi Let us mention now that matrices $A$ and $C$ are symmetric
($A^{\tau} = A$ and $C^{\tau} = C$), since the matrices $\al$ and
$\gam$ are symmetric. Also, if the Lagrangian $L(\dot{x},x,t)$ is
nonsingular ($\det {\al}\neq 0$), then the Hamiltonian $H(p,x,t)$
is also nonsingular ($\det {\bf A}\neq 0$). \vspace{2mm}

\noi The above calculations can be considered as a map from the
space of quadratic nonsingular Lagrangians ${\cal L}$ to the
corresponding space of quadratic nonsingular Hamiltonians ${\cal
H}\,.$ More precisely, we have $ \varphi: {\cal L}\longrightarrow
{\cal H}, $ given by \bea \nn
\varphi(L({\al},{\bet},{\gam},{\delta},{\xi} ,{\phi},{\dot x},x))
& = & H (\varphi_1(L),\varphi_2(L),
\varphi_3(L),\varphi_4(L),\vspace{3mm}
\\ & & \label{lh}\hspace{5.5mm} \varphi_5(L),\varphi_6(L),\varphi_7(L),\varphi_8(L))
\vspace{4mm}\\ & = & \nn H({A},{B},{C},{D},{E},{F},{p},{x})\,.\eea
From relation (\ref{clh}) it is clear that inverse of $\varphi$ is
given by the same relations (\ref{lcl}). This fact implies that
$\varphi$ is essentially involution, i.e. $\varphi\circ\varphi=
id\,.$


\subsection{Noncommutative case}
\noi In the case of noncommutative coordinates $[\,\hat x_k  ,
\hat x_j \, ] = i\, \hbar\,\theta _{kj}$, one can replace these
coordinates using the following ansatz (\ref{7}), \beq\label{ncc}
{{\hat x}}= {{\hat q}} - \frac12\,\, {\Theta}\, {{\hat p}}\,,\eeq
where ${{\hat x}}=(\hat x_1,\hat x_2,\hat x_3), \,\,{{\hat
q}}=(\hat q_1,\hat q_2,\hat q_3 ),\,\,{{\hat p}}=(\hat p_1,\hat
p_2,\hat p_3 )$ and \bea\nn  { \Theta} = \left(
\begin{array}{ccc}
0             & \theta _{12}  & \theta _{13}\\
-\,\theta _{12} & 0             & \theta _{23} \\
-\,\theta _{13} & -\,\theta _{23} & 0 \end{array} \right) . \eea
Now, one can easily check that $\hat q_i$ for $i = \ 1,2,3 \ $ are
mutually commutative operators (but do not commute with operators
of momenta,  $[\,\hat q_k , \hat p _j\, ] = i\,\hbar\, \delta
_{kj}$). \vspace{1mm}

\noi If we start with quantization of the nonsingular quadratic
Hamiltonian given by (\ref{ch}), i.e., $\hat H={
H}({A},{B},{C},{D},{E},{F},{{\hat p}}, {\hat x})$ and then apply
the change of coordinates (\ref{ncc}), we will again obtain
quadratic quantum Hamiltonian, ${\hat H}_{\Theta}\,:$
\begin{equation}
\hat H_{\Theta} =\langle\, {A}_{\Theta}\, \hat {p}, \hat {p}\,
\rangle + \langle\, {B}_{\Theta}\, \hat {q}, \hat {p}\, \rangle
+\langle\, {C}_{\Theta}\, \hat {q}, \hat {q}\, \rangle + \langle\,
{D}_{\Theta} , \hat {p}\, \rangle + \langle\, {E}_{\Theta} , \hat
{q}\, \rangle + {F}_{\Theta} ,
\end{equation}
where
\bea\nn & & {A}_{\Theta} = ({A} - {1\over 2}\,\, { \Theta}\, {B}
-{1\over 4}\, {\Theta}\, {C}\, { \Theta})_{sym}, \ \ \ \ \
{B}_{\Theta} = {B} + {\Theta}\, {C} , \vp{2mm}\\
& & {C}_{\Theta} = {C}, \ \ \ \ {D}_{\Theta} = {D} + {1\over
2}\,\, {\Theta}\, {E},\ \ \ \ {E}_{\Theta} = {E}, \ \ \ \
{F}_{\Theta} = {F} ,\label{th} \eea and  $_{sym}$ denotes
symmetrization of the corresponding operator. Let us note that for
the nonsingular Hamiltonian $\hat H$ and for  sufficiently small
$\theta_{kj}$ the Hamiltonian $\hat H_{\Theta}$ is also
nonsingular.

\vp{2mm}

\noi In the process of calculating path integrals, we need
classical Lagrangians. It is clear that to an arbitrary quadratic
quantum  Hamiltonian we can  associate the classical one
replacing operators by the corresponding classical variables.
Then, by using equations $$ \dot{q}_k = \frac{\partial
H_\Theta}{\partial p_k} , \ \ \ k=1,2,3,
$$ from such Hamiltonian we can come back to the
corresponding Lagrangian $$L_\Theta (\dot{q},q,t) = \langle p,
\dot{q} \rangle - H_\Theta (p,q,t),
$$ where
$$ p = \frac{1}{2} \, A_\Theta^{-1}\, (\dot{q} - B_\Theta\, q - D_\Theta) $$
is replaced in $H_\Theta (p,q,t)$. In fact, our idea is to find
connection between Lagrangians of  noncommutative and the
corresponding commutative quantum mechanical systems (with $\theta
= 0 $). This implies to find the composition of the following
three maps: \beq \label{com}
L_{\Theta}=(\varphi\circ\psi\circ\varphi)(L), \eeq where
$L_{\Theta}=\varphi(H_{\Theta}),\, H_{\Theta}=\psi(H)$ and
$H=\varphi(L)\,$ (here we use facts that $\varphi$ is an
involution given by formulas (\ref{lcl}), and $\psi$ is given by
(\ref{th})). More precisely, if \bea \nn & & L(\dot{x},x,t)
=\langle\, {\alpha}\, {\dot x}, {\dot x} \,\rangle + \langle\,
{\beta}\, {x}, {\dot x}\, \rangle +\langle\, {\gamma}\, {x}, {x
}\, \rangle + \langle\, {\delta} , {\dot x }\, \rangle + \langle\,
{\xi} , {x}\, \rangle + \phi ,\eea and \bea  & & L_{\Theta}
(\dot{q},q,t) =\langle\, {\alpha}_{\Theta}\, {\dot q}, {\dot q}
\,\rangle + \langle\, {\beta}_{\Theta}\, {q}, {\dot q}\, \rangle
+\langle\, {\gamma}_{\Theta}\, {q}, {q }\, \rangle + \langle\,
{\delta}_{\Theta} , {\dot q }\, \rangle + \langle\, {\xi}_{\Theta}
, {q} \, \rangle + {\phi}_{\Theta}\nn\, ,\eea then the connections
are given by \bea\nn  & & \hspace{-5mm} {\al}_{\Theta}= (
{\al}^{-\,1} - \frac 12 \,\, ({\bet}^{\,\tau}\, {\al}^{-\,1}\,
\Theta- \Theta\, {\al}^{-\,1}\, {\bet}) + \Theta\, {\gam}\, \Theta
- \frac 14\,\, \Theta\, {\bet}^{\tau}\, {\al }^{-\,1}\, {\bet}
\,\Theta )^{-\,1}\,, \vp{2mm} \\
 \nn & & \hspace{-5mm}  {\bet}_{\Theta}= {\al}_{\Theta}\, ( {\al}^{-\,1}\,
 {\bet} - \frac 12 \,\, \Theta\, {\bet}^{\,\tau}\, {\al}^{-\,1}\,
{\bet} + 2\, \Theta\, {\gam} )\,, \vp{2mm} \\
\nn & & \hspace{-5mm} {\gam}_{\Theta}=\frac 14\,\,
({\bet}^{\,\tau}\, {\al}^{-\,1}+\frac 12\,\, {\bet}^{\,\tau}\,
{\al}^{-\,1}\, \bet\,\Theta\, - 2\,  {\gam}\, \Theta )\,
{\al}_{\Theta}\, ({\al}^{-\,1}\,{\bet}- \frac 12\,\, \Theta\,
{\bet}^{\,\tau}\,
{\al}^{-\,1}\, {\bet}  \vp{2mm} \\
& & \hp{5mm} +\ 2\, \Theta\, {\gam}) -\frac 14\,\, {\bet
}^{\,\tau}\, {\al}^{-\,1}\, {\bet} +{\gam}\,, \lb{e1}\vp{2mm} \\
& & \hspace{-5mm}
{\delta}_{\Theta}={\al}_{\Theta}\,({\al}^{-\,1}\,{\delta}-\frac
12\,\,\Theta\,{\bet}^{\,\tau}\,
{\al}^{-\,1}\,{\delta} +\Theta\,\xi )\,, \nn \vp{2mm} \\
\nn & & \hspace{-5mm}  {\xi }_{\Theta} = \frac 12\, \,
(\bet^{\,\tau}\, {\al}^{-\,1}+\frac 12\, {\bet}^{\,\tau}\,
{\al}^{-\,1}\, \bet\,\Theta\, - 2\,  {\gam}\, \Theta )\,
{\al}_{\Theta}\, ({\al}^{-\,1}\,{\delta}
\vp{2mm} \\
\nn & & \hp{5mm} -\ \frac 12\,\, \Theta\, {\bet}^{\,\tau}\,
{\al}^{-\,1}\,{\delta} +  \Theta\, {\xi})-\frac 12\,\,
{\bet}^{\,\tau}\,{\al}^{-\,1}\, {\delta} + {\xi}\,, \vp{2mm} \\
\nn & & \hspace{-5mm}  {\phi }_{\Theta} = \frac 14\,\, \langle\,
{\delta}_{\Theta}\,, {\al}^{-\,1}\,{\delta}-\frac
12\,\,\Theta\,{\bet}^{\,\tau}\, {\al}^{-\,1}\,{\delta}
+\Theta\,{\xi}\,\rangle - \frac 14\,\, \langle\,
{\al}^{-\,1}\,{\delta}\,,{\delta}\,\rangle + {\phi}\,. \eea
\vp{3mm}

\noi It is clear that formulas (\ref{e1}) are very complicated and
that to find explicit  exact relations between elements of
matrices in general case is a very hard task. However, the
relations (\ref{e1}) are quite useful in all particular cases.

\subsection{Linearization and  some examples}

 \noi In this section we  try to introduce
 the linear change of coordinates to gain $L_\Theta (\dot{q},q,t)$ directly from
 $L(\dot{x},x,t)$ for some simple examples.
 Let $L(\dot{x},x,t)$ be a nonsingular quadratic Lagrangian given by (\ref{lg}).
 If we make the following change of variables:
  \vp{2mm} \beq\nn
\dot{x}=\dot{q}\,, \quad {x}=U(\Theta)\,\dot{q} +V(\Theta)\,{q}
+W(\Theta)=U\,\dot{q} +V\,{q} +W\,, \lb{e2}\eeq

\noi  we  obtain again quadratic Lagrangian
$L_{\Theta}(\dot{q},q,t),$ where \bea & &\hspace{-12mm} \nn
{\al}_{\Theta}= ( {\al} + U^{\,\tau}\, {\gam}\, U + {\bet}\,
U)_{sym}\,, \hspace{1.8cm}
{\bet}_{\Theta}=2\, U^{\,\tau}\, {\gam}\, V + {\bet}\, V, \vp{3mm} \\
\label{e3} & & \hspace{-12mm} {\delta}_{\Theta}={\bet}\, W + 2\,
U^{\,\tau}\, {\gam}\, W + {\delta} + U^{\,\tau}\, {\xi}\,,
\hspace{1cm} {\gam}_{\Theta}= V^{\,\tau}\,
{\gam} \, V,\vp{3mm} \\
\nn & & \hspace{-12mm}{\xi}_{\Theta} = 2\, V^{\,\tau}\, {\gam}\, W
+ V^{\,\tau}\,{\xi}, \hspace{3.02cm} {\phi}_{\Theta} = \langle\,
{\gam}\,W+ {\xi}, W \,\rangle + {\phi}\, , \eea and $U,\, V,\, W$
are matrices.

\noi It is clear that formulas (\ref{e3}) are  simpler than
(\ref{e1}). We will see that for some systems, by this linear
change of the coordinates, it is possible to obtain  connection
between Lagrangians $L(\dot{x},x,t)$ and $L_{\Theta}(\dot{q},q,t)$
\vspace{5mm}

\noi {\bf Example 1\,.} {\it Harmonic oscillator}, $n=2.$ In this
case we have \bea & & \hspace{-1cm} {\al}=\frac m2\,\, {\mathsl
Id},\quad {\bet}=0,\quad {\gam}=k\,{\mathsl Id},\quad \Theta =
\theta\, J, \label{e4} \eea  where $k<0$ is related to an ordinary
harmonic oscillator, \bea \nn & & \hspace{-1cm} J=\left( \ba{rc} 0 & 1\\
-\,1 & 0\ea \right),\ \ J^2=-\, {\mathsl Id},\quad {\delta}=0,
\quad {\xi}=(\xi_1^{},\xi_2^{}),\quad {\phi}=0\, , \eea

\noi and ${\mathsl Id}$ is $2\times 2$ unit matrix.

\noi Using formulas (\ref{e1}), one can easily find \bea
{\al}_{\Theta}=\frac {m}{2-k\,m\,\theta^2}\, {\mathsl Id},\qquad
 \label{e5} {\bet}_{\Theta}=\frac {2\,k\, m\,\theta}{2-k\,m\,\theta^2}\,
 J,\qquad {\gam}_{\Theta}=\frac {2\, k}{2-k\,m\,\theta^2}\,
{\mathsl Id}\,.\vp{2mm}\eea  From (\ref{e3}) it follows \bea\nn &
& {\al}_{\Theta}={\al}+U^{\tau}\,{\gam}\,U = {\al}+k\,
U^{\tau}\,U,\quad \mbox{and consequently}
\vp{3mm}\\
 & & {\al}_{\Theta}-{\al} =\frac {k\,m^2\,
 \theta^2}{2\,(2-k\,m\,\theta^2)}\,\,{\mathsl Id} .
\label{e6}\eea So, it is clear that matrix $U$ is proportional to
an orthogonal operator, i.e, $U=\l\, \tilde{U}.$ Now, from
(\ref{e6}) we obtain \bea\label{e7} \l=\varepsilon_{\l}\, \frac
{m\,\theta}{\sqrt{2}}\, \frac {1}{\sqrt{2-k\,m\,\theta^2}}\,,
\quad \mbox{where $\varepsilon_{\l}=\pm\, 1\,$ and
$\,2-k\,m\,\theta^2 > 0 \,.$} \eea Similarly, from (\ref{e5}) and
${\gam}_{\Theta}= V^{\,\tau}\, {\gam} \, V,$ we find that matrix
$V$ is also proportional to an orthogonal operator $V=\mu\,
\tilde{V},$ where \bea\label{e8} \mu = \varepsilon_{\mu}\,
\frac{\sqrt{2}}{\sqrt{2-k\,m\,\theta^2}}\,,\quad \mbox{
$\varepsilon_{\mu }=\pm\, 1\,$ and $\,2-k\,m\,\theta^2>0\,.$} \eea
It is known that \begin{footnotesize}\beq \label{e9} \ba{lll}
\tilde{U}=R(\psi)=\left( \ba{cr} \cos\,\psi & -\, \sin\,\psi \\
\sin\,\psi & \cos\,\psi\ea\right) & \mbox{or} &
\tilde{U}=R_1(\psi)=\left( \ba{rc}
-\,\cos\,\psi & \sin\,\psi \\
\sin\,\psi & \cos\,\psi\ea\right) \,,  \vp{4mm}\\
\tilde{V}=R(\omega)=\left( \ba{cr} \cos\,\omega & -\, \sin\,\omega \\
\sin\,\omega & \cos\,\omega\ea\right) & \mbox{or} &
\tilde{V}=R_1(\omega)=\left( \ba{rc}
-\,\cos\,\omega & \sin\,\omega \\
\sin\,\omega & \cos\,\omega\ea\right) \,,\ea\eeq
\end{footnotesize} where $0\leq \psi,\omega<2\, \pi\,.$  From (\ref{e3})
and (\ref{e5}), we find \bea & & \nn {\bet}_{\Theta}=2\,
U^{\,\tau}\, {\gam} \, V =
\frac{2\,k\,m\,\theta}{2-k\,m\,\theta^2}\,\, J, \eea

\noi and using  (\ref{e7})-(\ref{e9}), we have \bea
\tilde{U}^{\,\tau}\, \tilde{V}=\ve_{\l}\,\ve_{\mu}\, J\,.
\label{e10} \eea From (\ref{e10}), depending on
$\ve_{\l}\,\ve_{\mu}$ (1 or -\, 1), we have \bea\lb{e11}
\omega=\psi-\frac{\pi}{2} \qquad\mbox{or}\qquad
\omega=\psi+\frac{\pi}{2}\,.\eea If ${\bf\xi}=0,$  from the
relation ${\bf\phi}_{\Theta}=0,$ we obtain that $W=0\,.$ It
implies that in this case the transition from Lagrangian
$L(\dot{x},x,t)$ to the Lagrangian $L_{\Theta}(\dot{q},q,t)$ is
given by a linear change of the coordinates. \vp{3mm}

\noi Finally, let us show that in the case ${\bf\xi}\neq 0,$ it is
not  possible to obtain the transition from Lagrangian
$L(\dot{x},x,t)$ to the Lagrangian $L_{\Theta}(\dot{q},q,t)$  by a
linear change of the coordinates (\ref{e2}). From (\ref{e1}) and
(\ref{e3}),
we have \bea\nn & &
{\delta}_{\Theta}=2\,U^{\,\tau}\,{\gam}\, W+U^{\,\tau}\,{\bf\xi}
\eea
and consequently \bea & & W=\frac{1}{2\,\l\,k}\,\,\tilde{U}\,
{\delta}_{\Theta} - \frac{1}{2\,k}\,{\xi}\,.\lb{e12} \eea
Similarly, from the relation
${\xi}_{\Theta}=2\,V^{\,\tau}\,{\gam}\, W+V^{\,\tau}\,{\xi}\,,$
one can find
\bea& & W=\frac{1}{2\,\mu\,k}\,\,\tilde{V}\, {\xi}_{\Theta} -
\frac{1}{2\,k}\,{\xi}\,.\lb{e12a} \eea The relations (\ref{e12})
and (\ref{e12a}) imply \beq\lb{e13}
{\delta}_{\Theta}=-\,\frac{\l}{\mu}\,\,\tilde{U}^{\tau}\,\tilde{V}\,{\xi}_{\Theta}
= -\,\frac{|\l |}{|\mu |}\,\, J\, {\xi}_{\Theta}\,,\eeq From the
other side, the relations (\ref{e1}) and ({\ref{e5}) give \bea
{\delta}_{\Theta}=\frac{m\,\theta}{2-k\,m\,\theta^2}\,\, J\,
{\xi}\,, \lb{e14} \qquad
{\xi}_{\Theta}=2\,\frac{1-k\,m\,\theta^2}{2-k\,m\,\theta^2}\,\,
{\xi}\,.\eea Combining the relations (\ref{e13})  and
({\ref{e14}), we have\bea \nn & & \hspace{-1cm}
-\,\frac{1}{2-k\,m\,\theta^2} =
\frac{1-k\,m\,\theta^2}{2-k\,m\,\theta^2,}\quad \mbox{and
consequently $2-k\,m\,\theta^2=0\, ,$}\eea but it is impossible,
since  $\, 2-km\theta^2>0 $, according to (\ref{e7}) and
(\ref{e8}). \vp{5mm}

\noi In the first case (${\xi=0}$), the corresponding Lagrangian
$L_{\Theta}(\dot{q},q,t)$ is \bea\lb{e15} & & \hspace{-1cm}
L_{\Theta} = \langle\,{\al}_{\Theta}\,\dot{q},\dot{ q}\,\rangle +
\langle\,{\bet}_{\Theta}\, {q},\dot{ q}\,\rangle
+\langle\,{\gam}_{\Theta}\,{ q},{ q}\,\rangle+
\langle\,{\delta}_{\Theta},\dot{q}\,\rangle
+\langle\,{\xi}_{\Theta},{ q}\,\rangle +
{\phi}_{\Theta}\vp{2mm} \\
& & \nn \hp{-5mm} = \frac{1}{2-k\,m\,\theta^2} \, [ m\,(
{\dot{q}_1}^2 +{\dot{q}_2}^2)+ 2\,k\, ({q_1}^2 +{q_2}^2)+
2\,k\,m\,\theta\,( \dot{q}_1\,q_2 - q_1\,\dot{q}_2 )] \,. \eea
From (\ref{e15}), we obtain the  Euler-Lagrange equations,
\beq\lb{e15b} m\, \ddot{q}_1+2\, m\,k\,\theta\,\dot{q}_2-
2\,k\,q_1 =0 \qquad\mbox{and} \qquad m\, \ddot{q}_2 -2\,
m\,k\,\theta\,\dot{q}_1 - 2\,k\,q_2 =0\,. \eeq Let us remark that
the Euler-Lagrange equations (\ref{e15b}) form a coupled system of
second order differential equations, which is more complicated
than in commutative case ($\theta=0$). \vp{5mm}

\noi {\bf Example 2\,.} {\it A particle in a constant field},
$n=2\,.$

\noi This example is defined by the following data: \bea &
&\hspace{-10mm} {\al}=\frac m2\,\, {\mathsl Id},\quad
{\bet}=0,\quad {\gam}=0, \label{e16} \quad {\delta}=0, \quad
{\xi}=(\xi_1^{},\xi_2^{}),\quad {\phi}=0. \eea \

\noi Using the general composition formula (\ref{e1}), one can
easily find \bea \lb{e17} & & \hspace{-7mm} {\al}_{\Theta}=\frac
{m}{2}\,\, {\mathsl Id},\qquad\ \
 {\bet}_{\Theta}=0, \qquad\ \  {\gam}_{\Theta}=0, \qquad \ \
 {\xi}_{\Theta}={\xi}, \vp{3mm} \\
& & \hspace{-7mm} \nn {\delta}_{\Theta}=  \frac{m\,\theta}{2}\,
J\,{\xi}\, \qquad\ \ \mbox{and}\ \ \qquad
\phi_{\theta}=\frac{m\,\theta^2}{8}\,\langle\,{\xi},{\xi}\,\rangle\,,\eea
where $J$ is the same as in the first example (see (\ref{e4})).

\noi Now, using the method of the linear change of  coordinates,
from the relation (\ref{e3}), we have \bea\nn & & \hspace{-7mm}
{\al}_{\Theta}=\frac m2 \,\, {\mathsl Id},\qquad \ \
{\bet}_{\Theta}=0,\qquad\ \ {\gam}_{\Theta}=0,
\ \ \qquad {\xi}_{\Theta}=V^{\tau}\, {\xi}\,, \vp{3mm}\\
 & & \hspace{-7mm} {\delta}_{\Theta} =
  U^{\tau}\,\xi, \qquad\ \ \mbox{and}\ \ \qquad {\phi}_{\Theta}=
 \langle\,{\xi},W\,\rangle\, .
\label{e17a}\eea \noi From (\ref{e17}) and (\ref{e17a}), it is
easy to see that for \bea\lb{e17b}
U=-\,\frac{m\,\theta}{2}\,\,J\,,\qquad V={\mathsl Id}\qquad
\mbox{and}\qquad W=\frac{m\,\theta^2}{8}\, \xi\,,\eea the
linearization gives the same as the general formula. \vp{5mm}

\noi In this case, it is easy to find the classical action. The
Lagrangian $L_{\Theta}(\dot{q},q,t)$ is \bea\lb{e17c} L_{\Theta} &
= &
\frac{m}{2} \, \, ( {\dot{q}_1}^2 +{\dot{q}_2}^2 - \theta
\,(\xi_1\,
\dot{q}_2 - \xi_2\, \dot{q}_1 )) + \xi_1\, q_1 \vspace{2mm} \\
\nn &  & + \ \xi_2\, q_2
+\frac{m\,\theta^2}{8}\,({\xi_1}^2+{\xi_2}^2)\,. \eea The
Lagrangian given by (\ref{e17c}) implies the  Euler-Lagrange
equations, \beq\lb{e17d} m\, \ddot{q}_1 =\xi_1 \qquad\mbox{and}
\qquad m\, \ddot{q}_2 =\xi_2\,. \eeq \vp{2mm} Their solutions are:
\bea \lb{e17d} q_1(t)=
\frac{\xi_1\,t^2}{2\,m}+t\,C_2+C_1\qquad\mbox{and}\qquad q_2(t)=
\frac{\xi_2\,t^2}{2\,m}+t\,D_2+D_1,\eea  where $C_1,C_2,D_1$ and
$D_2$ are constants which have to be determined from  conditions:
 \bea\lb{e18} q_1(0)=x_1',\quad q_1(T)=x_1'' \quad
q_2(0)=x_2',\quad q_2(T)=x_2''.\eea  After finding the
corresponding constants, we  have
\bea & & q_j(t)= x_j'+ \frac{\xi_j\,t^2}{2\,m}+
t\,\left(\frac{1}{T}\,(x_j''-x_j') - \frac{\xi_j\,
T}{2\,m}\right)\,,\qquad j=1,2\,,
 \lb{e20} \vspace{2mm} \\
 \lb{e20a} & & \dot{q}_j(t)=
\frac{\xi_j\,t}{m}+ \frac{1}{T}\,(x_j''-x_j') - \frac{\xi_j\,
T}{2\,m}\,,\qquad j=1,2\,.
\eea
Using (\ref{e20}) and (\ref{e20a}),  we finally calculate the
corresponding action \bea & & \nn {\bar S}_\Theta
(x'',T;x',0)=\int\limits_{0}^T \, L_{\Theta}(\dot{q},q,t)\, d\,t =
\frac{m}{2\,T} \, [(x_1''-x_1')^2+(x_2''-x_2')^2]  \vp{3mm} \\ & &
\hp{5mm} +\ \frac{T}{2}\,\, [\xi_1\,(x_1''+
x_1')+\xi_2\,(x_2''+x_2')] \nn  - \frac{m\,\theta}2\,\,
[\xi_1\,(x_2''- x_2')  \vp{3mm} \\
& & \hp{5mm} -\ \xi_2\,(x_1''- x_1')]  - \frac{T^3}{24\, m}\,\,
({\xi_1}^2+{\xi_2}^2) - \frac{m\,T\, \theta}{8}\,\,
({\xi_1}^2+{\xi_2}^2)\,. \eea

\section{Concluding Remarks}

\noi Note that almost all results obtained in Section $2$ for
three-dimensional case allow straightforward generalization to any
$n\geq 2$ spatial dimensions. \vspace{1mm}

\noi  According to the formula (\ref{5}), the propagator in NCQM
with quadratic Lagrangians can be easily written down  when we
have the corresponding classical action ${\bar S}_\Theta
(x'',t'';x',t') = \int_{t'}^{t''} L_\Theta (\dot{q},q,t) d\, t ,$
where $q=q(t)$ is solution of the Euler-Lagrange equations of
motion. As a simple example we calculated ${\bar S}_\Theta
(x'',T;x',0)$ for a noncommutative regime of a particle on plane
in a constant field. \vspace{1mm}

\noi  Note that the path integral approach to NCQM has been
considered in the context of the Aharonov-Bohm effect \cite{7},
Aharonov-Bohm and Casimir effects \cite{8},  a quantum system in a
rotating frame \cite{9}, and the Hall effect \cite{turci} (see
also approaches \cite{mangano} and \cite{acatrinei}). Our approach
includes all systems with quadratic Lagrangians (Hamiltonians) and
some new results on path integrals on noncommutative spaces will
be presented elsewhere.
 \vspace{5mm}

\noi  {\bf Acknowledgements.} The authors thank S. Prvanovi\'c for
discussions. The work on this article was partially supported by
the Serbian Ministry of Science, Technologies and Development
under contract No 1426. The work of B.D. was also supported in
part by RFFI grant 02-01-01084.


\begin{thebibliography}{99} 

\bibitem{1} M.R. Douglas and N.A. Nekrasov, {\it Noncommutative Field Theory},
Rev. Mod. Phys. 73 (2001) 977-1025.

\bibitem{2} P. Aschieri, B. Jurco, P. Schupp, J. Wess, {\it Non-Commutative
GUTs, Standard Model and C,P,T}, Nucl. Phys. B651 (2003) 45-70,
arXiv: hep-th/0205214.

\bibitem{3} S. Bellucci, {\it Constant magnetic field and $2d$
non-commutative inverted oscillator}, arXiv: hep-th/0301227.

\bibitem{4} R.P. Feynman and A.R. Hibbs, {\it Quantum Mechanics and
Path Integrals}, McGraw-Hill Book Company, New York, 1965.

\bibitem{5} N. Moeller and B. Zwiebach, {\it Dynamics with Infinitely
Many Time Derivatives and Rolling Tachyons},  arXiv:
hep-th/0207107.

\bibitem{6}  Ya. Volovich, {\it Numerical Study of
Nonlinear Equations with Infinite Number of Derivatives}, arXiv:
math-ph/0301028.

\bibitem{lukierski} J. Lukierski, P.C. Stichel and W.J.
Zakrzewski, {\it Galilean-Invariant (2+1)-Dimensional Models with
a Chern-Simons-Like Term and D=2 Noncommutative Geometry}, Annals
of Phys. 260 (1997) 224-249, arXiv: hep-th/9612017.

\bibitem{7} M. Chaichian, A. Demichev, P. Pre\v snajder, M.M.
Sheikh-Jabbari and A. Tureanu, {\it Aharonov-Bohm Effect in
Noncommutative Spaces}, Phys. Lett. B527 (2002) 149-154, arXiv:
hep-th/0012175.

\bibitem{8} M. Chaichian, A. Demichev, P. Pre\v snajder, M.M.
Sheikh-Jabbari and A. Tureanu, {\it Quantum Theories on
Noncommutative Spaces with Nontrivial Topology: Aharonov-Bohm and
Casimir Effects}, Nucl. Phys. B611 (2001) 383-402, arXiv:
hep-th/0101209.

\bibitem{9} H.R. Christiansen and F.A. Schaposnik, {\it  Noncommutative
Quantum Mechanics and Rotating Frames}, Phys. Rev. D65 (2002)
086005, arXiv: hep-th/0106181.

\bibitem{turci} O.F. Dayi and A. Jellal, {\it Hall Effect in Noncommutative
Coordinates}, J. Math. Phys. 43 (2002) 4592, arXiv:
hep-th/0111267.

\bibitem{mangano} G. Mangano, {\it Path Integral Approach to Noncommutative
Spacetimes}, J. Math. Phys. 39 (1998) 2589-2591, arXiv:
gr-qc/9705040.

\bibitem{acatrinei} C. Acatrinei, {\it Path Integral Formulation of Noncommutative
Quantum Mechanics}, JHEP 0109 (2001) 007, arXiv: hep-th/0107078.



\end{thebibliography}
\end{document}